\documentclass[5p,twocolumn,sort&compress]{elsarticle}
\usepackage{graphicx}
\usepackage{amsfonts}
\usepackage{amssymb}
\usepackage{array}
\usepackage{amsmath}
\usepackage{verbatim} 
\usepackage{hyperref}
\usepackage{color}

\newcommand{\ket}[1]{\left\vert#1\right\rangle}
\newcommand{\bra}[1]{\left\langle#1\right\vert}
\newcommand{\pjct}[2]{\left\vert#1\right\rangle\!\left\langle#2\right\vert}
\newcommand{\tn}[1]{\textnormal{#1}}
\newcommand{\abs}[1]{\left|#1\right|}

\def\bra#1{\langle #1|}
\def\ket#1{|#1 \rangle}

\def\Tr{\mbox{Tr}}

\begin{document}

\title{
Characterizations and Quantifications  of Macroscopic Quantumness and Its Implementations using Optical Fields
}

\author{Hyunseok Jeong}
\author{Minsu Kang}
\author{Hyukjoon Kwon}

\address{Center for Macroscopic Quantum Control, 
Department of Physics and Astronomy,
Seoul National University, Seoul, 151-742, Korea}  

\date{\today}

\begin{abstract}
We present a review and discussions on characterizations and quantifications of macroscopic quantum states 
as well as their implementations and applications in optical systems. We compare and criticize different measures proposed to define and quantify macroscopic quantum superpositions and
extend such comparisons to several types of optical quantum states actively considered for experimental implementations within recent research topics. 
\end{abstract}

\maketitle

\section{Introduction}

Many quantum phenomena, often radically different from our intuitive predictions, are attributed to the fundamental principle of quantum superposition that a physical system can be in a linear superposition of two
distinct states.
When the principle of quantum superposition is applied to multipartite physical systems, it manifests another interesting feature of quantum mechanics, namely quantum entanglement.
Quantum entanglement is nonclassical correlation between local systems and 
is now widely referred to as a key resource for quantum information processing.
Quantum phenomena are observed typically on microscopic scales.
However, as illustrated in Schr\"{o}dinger's  famous cat paradox \cite{Schrodinger1935},
quantum mechanics does not preclude, in principle, the possibility of a macroscopic object being in a quantum superposition or
being a part of quantum entanglement.
Natural questions then follow.
If ``macroscopic {\it and} quantum'' is somehow a possible combination,
how can we define, characterize and quantify ``macroscopic quantumness'' or ``quantum macroscopicity''?
Further, how and to what extent can we implement such macroscopic quantumness?

Of course, scientists have tried to answer these questions.
In the early days of quantum mechanics,
such attempts were primarily in the area of interpretations or philosophical discussions, as they are not within the reach
of experimental tests or implementations.
Since then, however, a remarkable development of quantum and atom optics has been brought about, which has paved a way to control and detect individual quantum systems at the level of single photons and atoms. Based on this progress, further efforts are being made to collectively control larger quantum systems, that is also closely related to the ability to perform quantum information processing \cite{Div}. 
Now, we may say that a significant amount of efforts made in physics and optics research in the last few decades are  
more or less
 related to explorations of macroscopic quantumness.
In this paper, we review two major research topics on macroscopic quantumness --
its quantifications and physical implementations using optical fields  -- and attempt to make remarks on them.

\section{Characterization and quantification of macroscopic quantumness}

We want to know whether and/or how much a physical system is {\it both} macroscopic {\it and} quantum.
Needless to say, it is not sufficient for a state to be either macroscopic or quantum.
It should be {\it quantum-mechanically macroscopic} or {\it macroscopically quantum}.
This point may sound too obvious to make, but it is nontrivial to technically define and quantify  macroscopic quantumness.
Leggett posed a question along this line as ``What is the correct measure of `Schr\"odinger's-cattiness'?'' 
and commented, ``Ideally, one would like a quantitative measure
which corresponds to our intuitive sense'' \cite{Leggett2002}.

A number of proposals have been made for quantification of macroscopic quantum superpositions \cite{Leggett1980,Dur,Shimizu2002,Bjork,Shimizu2005,Cavalcanti,Korsbakken,Mar,Korsbakken2,LeeJeong2011,Frowis2012NJP,Nim2013,Sekatski2014}
based on the effective number of particles that involve the superposition
\cite{Leggett1980,Dur,Mar,LeeJeong2011},
distinguishability between the constituent states 
\cite{Dur,Bjork,Cavalcanti,Korsbakken,LeeJeong2011,Sekatski2014} and operational interpretations \cite{Bjork,Frowis2012NJP,Nim2013}. 
In the context of this paper, we would like to 
pose three requirements for a desirable measure of macroscopic quantumness.
First, it should be applicable to a wide range of states, not limited to a specific type of states.
Second, we prefer to have a measure that quantifies the degree of a genuine superposition against a classical mixture,
together with its effective size factor.
In other words, it should be applicable not only to pure states but also to mixed states.
Third, if a state is given, the degree should be unambiguously determined.
These points are important because our motivation is to compare different types of states 
being considered as candidates for macroscopic quantum superpositions using a conclusive measure.
In view of these points, we shall review in chronological order such measures proposed to quantify
macroscopic quantumness.

\subsection{Disconnectivity}
In 1980, Leggett in his pioneering work defined a measure called ``disconnectivity" $D$  \cite{Leggett1980}  that
quantifies genuine multipartite quantum correlations.
Suppose that we are interested in characterizing an $N$-mode bosonic system $\rho_{N}$.
We first obtain a reduced density operator $\rho_n$ ($n<N$) from $\rho_N$ by tracing out every mode except for  $n$ arbitrarily chosen modes.
Quantity $\delta_n$ is introduced as
\begin{equation}
\delta_n = \frac{S_n}{\min_m(S_m+S_{n-m})} 
\end{equation}
where $S_n=-\Tr[\rho_n\ln\rho_n]$ is von-Neumann entropy of $\rho_n$. By definition, $\delta_n$ is set to be $1$ when both the numerator and the denominator are zero, and $\delta_1\equiv0$.
The disconnectivity $D$ is defined as the largest integer $n$ that makes $\delta_n$ the smallest.
For an ideal Greenberger-Horne-Zeilinger (GHZ) state $\propto\ket{\phi}^{\otimes N} +\ket{\phi^\bot}^{\otimes N}$ with an orthogonal basis $\{\ket{\phi}, \ket{\phi^\bot}\}$, it is clear that $S_N=0$ and $S_n=1$ for $n\neq N$. Its disconnectivity is then $D=N$ that is the largest $n$  minimizing $\delta_n$ to be $0$.
However,  $D$ is always $1$ for mixed states  $\propto \pjct{\phi}{\phi}^{\otimes N}\!\!+\pjct{\phi^\bot}{\phi^\bot}^{\otimes N}+\Gamma( \pjct{\phi}{\phi^\bot}^{\otimes N}+\tn{H.c.})$  regardless of the values of $N$ and $\Gamma$ as far as $\Gamma<1$. This means that  macroscopic quantumness for partially mixed states
cannot be identified by $D$.

Leggett pointed out that so called ``macroscopic quantum phenomena" such as superconductivity or superfluidity do not require the existence of a high-$D$ state.
Superfluidity can be explained by a product of identical bosonic states of which disconnectivity is obviously $1$.
A superconducting system described by $N$ Cooper pairs, $(\ket{\uparrow\uparrow}+e^{i\phi}\ket{\downarrow\downarrow})^{\otimes N}$, also shows a small value of disconnectivity $D=2$ regardless of $N$.
On the other hand,  multi-mode quantum correlations, in the form of pure states, always give large values of $D$.

Even though disconnectivity sensibly quantifies multipartite correlations for pure states, it is not always sensitive to distinguishability between the constituent  states.
This point shall be clearer in the following example discussed by D\"{u}r {\it et al.}~\cite{Dur}.

\subsection{Effective size of Greenberger-Horne-Zeilinger type states}
D\"{u}r {\it et al.}~\cite{Dur} 
considered the effective size of a  generalized form of the Greenberger-Horne-Zeilinger (GHZ) state:
\begin{equation}
\ket{\psi_\epsilon}=\frac{1}{\sqrt{K}}\left(\ket{0}^{\otimes N}+\ket{\epsilon}^{\otimes N}\right)
\label{eq:gen-ghz}
\end{equation}
where  $\ket{\epsilon}=\cos{\epsilon}\ket{0}+\sin{\epsilon}\ket{1}$.
The disconnectivity of this state is found to be $D=N$ regardless of $\epsilon$ even though
$\epsilon$ obviously contributes to the distinguishability between $\ket{0}^{\otimes N}$ and $\ket{1}^{\otimes N}$. It would be of  particular interest to know whether $\ket{\psi_\epsilon}$ can still be a macroscopic superposition in comparison to an ideal GHZ state when $\epsilon$ takes a small value.
To put it concretely, they attempt to find out what size of the ideal GHZ state (i.e. $\ket{\psi_{\epsilon=\pi/2}}$) state
$\ket{\psi_\epsilon}$ is equivalent to.
Two different methods lead to the same result that the effective size of state $\ket{\psi_\epsilon}$
as a macroscopic superposition approaches  $N\epsilon^2$ when $\epsilon\ll 1$.
The first method is based on the rate of decoherence and the second is related to the GHZ entanglement distillation by local operations and classical communication (LOCC).
Consider a dephasing process that is described by a  completely positive map 
\begin{equation}
\mathcal{E}(\rho)=p_0 \rho+(1-p_0)\sigma_z \rho \sigma_z
\end{equation}
where $p_0=(1+e^{-\gamma t})/2$, $t$ is time, $\gamma$ is the dephasing rate, and $\sigma_z$ is the Pauli-$z$ operator.
They show that 
the trace-norm of the off-diagonal elements for state $\ket{\psi_\epsilon}$
 is $ e^{-\gamma N \epsilon^2 t}$  when  $\epsilon\ll 1$ and that  it is $ e^{-\gamma N t}$
 for state $\ket{\psi_{\epsilon=\pi/2}}$.
Based on this comparison, they conclude that the effective size of the generalized GHZ state 
 $\ket{\psi_\epsilon}$ is equivalent to that of an ideal GHZ state  $\ket{\psi_{\epsilon=\pi/2}}$ of size $N\epsilon^2$ for small values of $\epsilon$.
This conclusion is also derived from depolarizing decoherence where it gives exactly the same decay rate. 
Of course,  
this analysis is limited  only to a very specific type of states in the form of Eq.~(\ref{eq:gen-ghz}).

\subsection{Interference-based measure}

Bj\"{o}rk and Mana's suggestion  \cite{Bjork} is based on their observation that 
a quantum superposition is more sensitive than its constituent states for
interferometric applications.
Let us consider a pure state $\ket{\psi}$ with a measurement outcome distribution of observable $\hat{A}$ as 
\begin{equation}
\abs{\langle A \vert \psi \rangle}^2= \abs{\psi(A)}^2=f(A-A_1)+f(A-A_2)
\end{equation}
where $f(A-A_i)$ is a function of a reasonably smooth form centered at $A_i$ with width $\Delta A$.
The operator $e^{i\theta \hat{A}}$ is applied to state $|\psi\rangle$
where $\theta$ corresponds to the degree of the interaction time and strength.
Generally, the overlap between the original and evolved states is
\begin{equation}
\abs{\bra{\psi}e^{i\theta \hat{A}}\ket{\psi}}=2\abs{\cos{\frac{\theta(A_2-A_1)}{2}}}\cdot\abs{\int\! dA \ e^{i\theta A} f(A)},
\label{eq:overlap}
\end{equation}
which becomes zero when $\theta=\theta_\tn{sup}\equiv \pi/(A_2-A_1)$.
In other words, the original state $\ket{\psi}$ evolves to a state that is orthogonal 
to the original one at a certain interaction time $\theta=\theta_\tn{sup}$.
Meanwhile, if there was only a single peak $f(A-A_i)$ ($i=1$ or $2$) for the distribution of $|\psi\rangle$, the overlap simply would be $\abs{\int\! dA \ e^{i\theta A} f(A)}$ regardless of $i$ and 
its first local minimum (or a half of the initial value) would be found at $\theta\approx\theta_{\tn{sing}}\equiv  \pi/\Delta A$.
The measure of a macroscopic superposition in terms of interferometric sensitivity is defined as the ratio of the two interaction times:
\begin{equation}
\frac{\theta_\tn{sing}}{\theta_\tn{sup}}=\frac{\abs{A_2-A_1}}{\Delta A}.
\end{equation}
Bj\"{o}rk and Mana's approach has a distinguishing feature as an operational measure based on a physical application even though it is devised only for pure states.
There is an ambiguity applying this measure to mixed states about how close the evolved state should be to the orthogonal state in determining $\theta_{\rm sup}$. 
This may lead to ambiguity in comparing different types of states.

\subsection{Indices $p$ and $q$ based on correlations of local observables}
Shimizu and Miyadera proposed index $p$  \cite{Shimizu2002}
that determines whether a  given form of $N$-mode state $|\psi_N\rangle$ becomes macroscopically quantum as $N$ increases. The index $p$ is obtained as
\begin{equation}
\max_{\hat{A}} V_\psi(\hat{A})=O(N^p),
\label{eq:pindex}
\end{equation}
where $V_\psi(\hat{A})$ is the variance
for observable $\hat A$ with state $|\psi\rangle$, $N$ is the number of modes, and $f(N)=O(N^p)$ if $\lim_{N\rightarrow \infty}f(N)/N^p$ is a nonzero constant.
The maximum in Eq.~(\ref{eq:pindex}) is taken over all possible additive observables represented by $\hat{A}=\sum_{n=1}^N \hat{A}_n$ where  $\hat{A}_n$ is a local observable for mode $n$. 
The value of  index $p$ is found to be $p=2$ for GHZ-type entanglement
while $p=1$ for a simple product form of state.
A state is considered to be macroscopically quantum if its index is $p=2$.
The index $p$, however,  cannot be applied to mixed states. For example, a mixed state $\pjct{0}{0}^{\otimes N}+\pjct{1}{1}^{\otimes N}$ have  the same value of index $p=2$ as
an ideal GHZ state.

Shimizu and Morimae generalized the measure for arbitrary mixed states \cite{Shimizu2005}.
The index $q$ for state $\rho$ is 
\begin{equation}
\max_{\hat{A},\hat{\eta}} (\langle \hat{C}_{\hat{A},\hat{\eta}} \rangle,N) =O(N^q)
\end{equation}
 where $\hat{\eta}$ is an arbitrary  projection operator satisfying $\hat{\eta}^2=\hat{\eta}$
 and $\hat{C}_{\hat{A},\hat{\eta}}= [ \hat{A},[\hat{A},\hat{\eta}]]$.
The correlation can be  represented as
$\langle \hat{C}_{\hat{A},\hat{\eta}} \rangle= \sum_{i,j} (a_i-a_j)^2 \langle a_i|\hat{\eta}|a_j\rangle\langle a_j | \rho | a_i \rangle$ where $|a_i\rangle$  is an eigenstate of $\hat A$ with eigenvalue $a_i$.
It becomes $O(N^2)$ so that $q=2$ when there is non-negligible $O(1)$ coherence $\langle a_j | \rho | a_i \rangle$
between the macroscopically distinct states of $\abs{a_i - a_j}=O(N)$.
The indices  $p$ and $q$ are equivalent for pure state, i.e., index $p$ is a special case of index $q$.

The index $q$, as well as index $p$,  does not give a value for a given {\it state}; rather it identifies what {\it kinds} (or forms) of states can scale to be macroscopically quantum when they become large multipartite states of $N\gg1$. Therefore, it cannot be directly applied to a single-mode state such as $|a_{k}\rangle+|a_{l}\rangle$ where its macroscopicity depends on $|a_k-a_l|$, nor is it a quantifier of macroscopic quantumness for a given state.

\subsection{Inequality for testing macroscopic superpositions}
Cavalcanti and Reid proposed an inequality of which violations verify macroscopic superpositions
of continuous-variable states \cite{Cavalcanti}.
 Suppose a generalized macroscopic superposition
\begin{equation}
c_+ \ket{\psi_+} + c_0 \ket{\psi_0}+ c_- \ket{\psi_-}
\label{eq:gs}
\end{equation}
and a pointer-measurement $\hat{X}$ giving macroscopically ranged outcomes $x$.
The domain for $x$ can be partitioned into three regions for $I=-1$, $0$, and $+1$ which correspond to $x\leq-S/2$,
 $-S/2< x < S/2$ and $x\geq S/2$, respectively.
If the reference value $S$ is sufficiently large, two region $I=-1$ and $+1$ are called macroscopically distinct.

In contrast to Eq.~(\ref{eq:gs}), a mixed state
\begin{equation}
\rho = p_L \rho_L + p_R \rho_R
\label{eq:rhoLR}
\end{equation}
is not a macroscopic superposition 
in the sense that the outcomes of $\rho_L$ only spread for $x<S/2$ $(I=-1,0)$ and those of $\rho_R$ for $x>-S/2$ $(I=0,+1)$. 
In other words, state (\ref{eq:rhoLR}) does not incorporate a macroscopic superposition since the coherence element is $\bra{\psi_+} \rho \ket{\psi_-}=0$.
The authors derived an inequality that should be satisfied by state~(\ref{eq:rhoLR}):
\begin{equation}
(\Delta^2_\tn{ave}x+P_0 \delta)\Delta^2p \geq 1
\label{eq:Cavalcanti}
\end{equation}
with 
\begin{equation}
\begin{aligned}
&\Delta^2_\tn{ave}x=P_+ \Delta^2_+ x + P_-\Delta^2_- x,\\
&\delta = \{ (\mu_+ + S/2)^2+(\mu_- - S/2)^2 + S/2\}+\Delta_+^2 x + \Delta^2_- x
\end{aligned}
\end{equation}
where
$P_\pm(x)$ are the normalized probability distributions for regions $I=\pm1$, and $\mu_\pm$ and $\Delta_\pm^2x$ are their means and variances, respectively.

Cavalcanti and Reid's inequality can be applied to a wide range of states compared to previous measures, i.e., applicable to arbitrary single-mode continuous-variable states and may be extended to multi-mode continuous-variable states if appropriate measurements are defined.
 However, it does not provide a degree of macroscopic quantumness, but it works as a criterion of macroscopic quantumness for a given state with respect to an arbitrarily chosen scale $S$.
Its calculations involve nontrivial numerical integrations and  optimization processes. 
Marquardt {\it et al.} experimentally demonstrated violations of this inequality using Gaussian states
and showed that it critically depends on purity of the states
  \cite{CavalcantiExp}.

\subsection{Measurement-based measure}

Korsbakken {\it et al}. \cite{Korsbakken} suggested an effective size of an $N$-particle superposition state in the form of $\ket{\psi_N}\propto \ket{A}+\ket{B}$ based on how many measurements are required to distinguish $\ket{A}$ and $\ket{B}$.
The size measure for state $|\psi_N\rangle$ is defined by
\begin{equation}
C_\delta(\ket{\psi_N})= \frac{N}{n_{\min}}
\end{equation}
for given probability $1-\delta$  $(\delta\ll1)$ of distinguishing the two constituent states by measuring $n_{\min}$ number of particles.
The discrimination probability is calculated as
\begin{equation}
P=\frac{1}{2}\left(\Tr[\rho_A^{(n)}E_A^{(n)}]+\Tr[\rho_B^{(n)}E_B^{(n)}]\right)
\end{equation}
where $\rho_{A,B}^{(n)}$ are reduced density matrices for $n$ particles ($n\leq N$) and $E_{A,B}^{(n)}$ 
are positive-operator valued measurements (POVMs) acting nontrivially only on $n$ particles.
The maximum value of the discrimination probability  is obtained when the POVM is a projective measurement in the eigenbasis of $\rho^{(n)}_A-\rho^{(n)}_B$ 
\begin{equation}
P=\frac{1}{2}+\frac{1}{4}\Tr | \rho^{(n)}_A-\rho^{(n)}_B | .
\end{equation}
Obviously, the two constituent states of an ideal GHZ state $|{\rm GHZ}_N\rangle=\ket{0}^{\otimes N}+\ket{1}^{\otimes N}$ can be distinguished by a single particle $\sigma_z$ measurement so that $C_\delta(\ket{{\rm GHZ}_N})=N$ regardless of $\delta$.

Korsbakken \textit{et al.}  applied this measure to superconducting flux qubits \cite{Korsbakken2} where a superposition of macroscopically distinct values of currents and magnetic moments is observed. They found the effective size of the flux qubits is surprisingly (but not trivially) small despite the apparent large difference in macroscopic observables. The reason is that only a small fraction of all electrons contribute to the superposition, while their speeds are high enough to yield large currents or magnetic moments.

There is an ambiguity in using this measure due to the arbitrariness of the choice of $\delta$.
For example, the size of the superposition in the form of $|{\rm D}_N\rangle\propto \ket{0}^{\otimes N}+\sum_{k=0}^N\ket{1}^{\otimes k}\ket{0}^{\otimes{N-K}}$ is found to be $C_\delta(\ket{{\rm D}_N})=2\delta(N+1)$ \cite{Frowis2012NJP}.
Therefore, for example, which of the two states, $\ket{{\rm GHZ}_N}$ and $\ket{{\rm D}_N}$, is more ``macroscopically quantum'' depends on the choice of $\delta$.

\subsection{Measure based on effective number of particles participating in the superposition}

Marquardt \textit{et al.} define the size of quantum superposition by a number of particles that effectively involve  the superposition \cite{Mar}.
More precisely, for given constituent states $\ket{A}$ and $\ket{B}$ they count how many single-particle operations are required to convert $\ket{A}$ into $\ket{B}$ or vice versa.
As a simple  example, an $N$-product horizontal-polarization state $\ket{H}^{\otimes N}$ can be converted into and a vertical-polarization state $\ket{V}^{\otimes N}$ by acting $\prod_{i=1}^N \hat{a}_{V,i}^\dagger \hat{a}_{H,i}$, 
where $\hat{a}^\dagger$ and  $\hat{a}$ are creation and annihilation operators for corresponding modes of the subscripts.
This operation is an $N$ number of single-particle operations, and the size of superposition $(\ket{H}^{\otimes N}+\ket{V}^{\otimes N})/\sqrt{2}$ is $N$.

In general, $\ket{B}$ is obtained by
superposing different states as $\ket{B}=\sum_{d=0}^N \beta_d \ket{\beta_d}$, where $\ket{\beta_d}$ is a state converted from $\ket{A}$ by at least $d$ single particle operations.
The average effective particle number is then $\sum_d\abs{\beta_d}^2d$.
In Ref.~\cite{Mar}, the authors considered $\ket{A}\propto(\hat{a}^\dagger)^N\ket{0}$ and $\ket{B}\propto(\cos\theta \hat{a}^\dagger+\sin\theta \hat{b}^\dagger)^N\ket{0}$ for modes $a$ and $b$.
The constituent state $\ket{B}$ is expanded as
\begin{align}
\ket{B}=\sum_{d=0}^{N}\frac{\beta_{d}(\theta) \ \hat{b}^{\dagger d} \hat{a}^{\dagger {N-d}} }{\sqrt{d!(N-d)!}} \ket{0}
\label{eq:Expand}
\end{align}
with coefficients $\beta_{d}(\theta)$ \cite{Mar}.
The state $\hat{b}^{\dagger d} \hat{a}^{\dagger {N-d}}\ket{0}$ can be obtained by applying  $d$ times of the single particle operation $\hat{b}^\dagger\hat{a}$ to $(\hat{a}^\dagger)^N\ket{0}$ implying that its effective number is $d$.
The average effective number is then  $N\sin^2\theta$ and
it becomes $N$ for an ideal GHZ state of $\theta=\pi/2$.
It is obvious that this measure cannot be applied to mixed states.

\subsection{Measure $\cal I$ based on the phase space structure}

In the phase space of mutually conjugate variables, the Wigner function of a macroscopic quantum superposition typically shows an interference pattern with a high frequency.
Taking note of this point, Lee and Jeong \cite{LeeJeong2011} defined a measure of macroscopic quantumness  for an arbitrary harmonic-oscillator state. It simultaneously quantifies two different kinds of essential information:
the degree of quantum coherence and the effective size of the physical system that involves the superposition.
The basic idea is to take an integral $\int \!\! d^2 \mathbf{\xi} \left( \xi_r^2 + \xi_i^2 \right) \left| \chi \left( \mathbf{\xi} \right) \right|^2$,
where $\chi(\xi) =
		\Tr [\, \rho\, e^{\xi\, \hat{a}^{\dagger} - \xi^{*} \hat{a}}\, ]$ is the characteristic function for state $\rho$,  in order to measure
``frequency'' and ``magnitude'' of the interference fringes at the same time in the Wigner representation.
In a continuous-variable phase space, it effectively quantifies {\it both} ``how widespread'' the constituent states of a given state are {\it and} ``how quantum mechanically pure'' coherences between those constituent states are, at the same time.

 The formal definition of the measure $\mathcal{I}$ for an $M$-mode harmonic oscillator system
\cite{LeeJeong2011} is only slightly different:
\begin{equation}
	\label{eq:integ}
\mathcal{I} \left( \rho \right)
= \frac{1}{2 \pi^M} \!\! \int \!\! d^2 \boldsymbol{\xi}
 \sum_{m = 1}^M \! \left[ \left| \xi_m \right|^2  - 1 \right] \left| 
 \chi \left( \xi_1,\, \xi_2,\, \cdots,\, \xi_M \right) \right|^2  
\end{equation}
where $\chi \left( \xi_1,\, \xi_2,\, \cdots,\, \xi_M \right) $ is the $M$-mode characteristic function, $\int d^2\boldsymbol{\xi} = \int d^2\xi_1 \int d^2\xi_2 \cdots \int d^2\xi_M$.
The value  $-1$ has been inserted to make any coherent states or their product states (regardless of their amplitudes) a reference with  $\mathcal{I}=0$, but it may be removed to guarantee positivity of the measure \cite{Gong,reply}.
The maximum possible value of $\cal I$ for an optical state is shown to be its average photon number
\cite{LeeJeong2011}.

Assuming the photon loss condition, 
$d\rho/d\tau = \hat{a} \rho \hat{a}^\dagger -\{ \hat{a}^\dagger \hat{a},\rho \}/2$ with $\tau$=(decay rate $\times$ time),
it turns out that the measure is equivalent to the purity decay rate of the state
\begin{equation}
\mathcal{I(\rho)} = -\frac{1}{2} \frac{d \mathcal{P}(\rho)}{d\tau}
\label{eq:ad}
\end{equation}
where $\mathcal{P}(\rho)=\Tr[\rho^2]$ is the purity of state $\rho$. 
This is consistent with the rapid decay of macroscopic superpositions, 
and the purity decay rate itself and Eq.~(\ref{eq:ad}) may be an alternative definition of $\cal I$.
It is possible to detect $\cal I$ for optical states using overlap measurements without full tomography of quantum states \cite{JeongDetect2014}.

The measure $\cal I$ is applicable to arbitrary harmonic oscillator systems including mixed and multi-mode states. It is decomposition-independent and straightforward to calculate for any states represented in the phase space, giving definite values for direct comparison between different types of states. 

\subsection{Fisher information as a measure of genuine macroscopic quantum effects}

Fr\"owis and D\"ur  suggested using the Fisher information as a measure of macroscopic quantumness for spin systems \cite{Frowis2012NJP}.
When estimating an unknown parameter, $\phi$, caused by an unitary evolution $e^{i \phi\hat{A} }$ with Hamiltonian $\hat{A}$, 
the classical limit of the estimation uncertainty is determined by $\Delta\phi \geq 1/\sqrt{N}$ where
$N$ is the system size. It is well known that this limit can be lowered down to $\Delta\phi \geq 1/N$ 
using a quantum mechanically correlated probe state. 
More generally, the limit of the estimation uncertainty
using probe state $\rho$
  is given by the Cram\'er-Rao bound \cite{Cramer}, $\Delta\phi \geq 1/\sqrt{F(\rho,\hat{A})}$,  where $F(\rho,\hat{A})$ 
is the quantum Fisher information\footnote{$F(\rho,\hat{A})=2\sum_{i,j=1}^{2^N} {(\pi_i-\pi_j)^2}/{(\pi_i+\pi_j)}|\bra{i}\hat{A} \ket{j}|^2$ where $\pi_i$ and $\ket{i}$ are eigenvalues and eigenvectors of $\rho$.}
  for a given additive Hamiltonian $\hat{A}$.
Therefore, $F(\rho,\hat{A})=O(N^2)$ implies that state $\rho$ is capable of the parameter estimation over the classical limit.
Noting that this improvement is attributed to long-range quantum correlations in the state $\rho$,
Fr$\ddot{\tn{o}}$wis and D$\ddot{\tn{u}}$r call it a genuine macroscopic quantum effect, not an accumulation of microscopic effects \cite{Frowis2012NJP}.
They call a quantum state $\rho$ macroscopic if
$\max_{\hat{A}\in \mathcal{A}}F(\rho,\hat{A})=O(N^2)$,
where the maximization is taken over the entire set $\mathcal{A}$ of additive operators.

However, there are some cases where individual particles do not show long-range quantum correlation while local groups of them do \cite{Frowis2012NJP}.
Therefore, 
the aforementioned measure needs to be generalized to
\begin{equation}
\max_{\hat{A}'\in \mathcal{A}'}F(\rho,\hat{A}')=O(n^2)
\label{eq:d2}
\end{equation}
where $\mathcal{A}'$ is a set of extended additive operator $\hat{A}'=\sum_{i=1}^n \hat{A}_i$ with each $\hat{A}_i$ locally acts on $n=O(N)$ distinct groups of particles sized $O(1)$.
As indices $p$ and $q$, Fr\"owis and D\"ur's approach in the forms of Eq.~(\ref{eq:d2}) cannot be directly applied to a single-mode state. Such a measure is defined as
\begin{equation}
N_\tn{eff}=\max_{{\hat A'}\in \mathcal{A'}}\Big\{\frac{F(\rho,\hat{A}')}{4n}\Big\}
\end{equation}
giving a definite number for a given state.

This proposal is applicable to arbitrary spin systems. It was applied to cloned quantum states to find that they are macroscopically quantum \cite{FrowisDurPRL2012}.
Fr\"owis and D\"ur also compared  \cite{Frowis2012NJP} several measures  for spin systems \cite{Dur,Shimizu2002,Bjork,Korsbakken,Mar} and concluded that index $p$  and their Fisher-information-base approach detect the most broad set of macroscopic quantum states among those.

\subsection{Measure based on minimal extension of quantum mechanics}

Nimmrichter and Hornberger  \cite{Nim2013} call a quantum superposition of a mechanical system to be macroscopic, if its experimental demonstration allows one to rule out even a minimal modification of quantum mechanics.
In order to specify such minimal modification, they consider an additional generator $\mathcal{L}_N$ to the von Neumann equation for $N$ particle density matrix $\rho_N$ as
\begin{equation}
\frac{\partial \rho_N}{\partial t} = \frac{1}{i\hbar}\left[ H, \rho_N \right] +\mathcal{L}_N (\rho_N),
\label{eq:vonNeumann}
\end{equation}
where 
\begin{align}
\mathcal{L}_N (\rho_N) =& \frac{1}{\tau_e}\int d^3s~ d^3q ~g_e(s,q)\Big[W_N({\bf s,q})\rho_N W_N^\dagger({\bf s,q})\nonumber\\
&-\left\{W_N^\dagger({\bf s,q})W_N({\bf s,q}),\rho_N\right\} \Big]
\end{align}
and $W_N({\bf s,q})=\sum_{n=1}^N \frac{m_n}{m_e} W_n(\frac{m_e}{m_n}{\bf s},{\bf q})$ is the weighted sum of the single particle operators $W_n({\bf x},{\bf p})=\exp[\frac{i}{\hbar}({\bf \hat{P}}\cdot{\bf x}-{\bf p}\cdot {\bf \hat{X}})]$ giving position translation ${\bf x}$ and momentum boost ${\bf p}$ of the $n$-th particle of mass $m_n$.
Here, $m_e$, $\tau_e$, and $g_e(s,q)$ are the mass, the coherence time parameter, and the normalized distribution function for the reference particle, respectively.
The reference particle is chosen to be an electron, and the distribution $g_e$ is taken to be a Gaussian distribution with standard deviations $\sigma_s$ and $\sigma_q$ for position and momentum respectively.
The role of the generator $\mathcal{L}(\rho_N)$ in Eq.~(\ref{eq:vonNeumann}) is to wipe out the coherence from the original distribution of $\rho_N$ in the position-momentum phase space.

For an experimental demonstration of a quantum superposition in a mechanical system, it rules out a certain region of the modification parameter so that there is a lower bound of the time parameter $\tau_e$.
If such a lower bound is larger, the superposition becomes more macroscopically quantum.
The lower bounds of $\tau_e$ for interference experiments with neutrons, electrons, Bose-Einstein condensates, and molecules are obtained  \cite{Nim2013,mechanical-review}, and then the measure of macroscopicity is defined as
\begin{equation}
\mu=\log_{10}\left(\frac{\tau_{e,\tn{max}}}{1~\tn{second}}\right),
\end{equation}
where $\tau_{e,\tn{max}}$ is the greatest lower bound of the time parameter $\tau_e$.
State-of-the-art interferometers achieve macroscopicities of up to $\mu\approx12$ \cite{Nim2013}.
It can be applied to any mechanical phenomena and successfully addressed various experiments  \cite{Nim2013,mechanical-review}.
However, it may require more investigations to find out whether macroscopicity witnessed by this measure is in line with the idea of genuine macroscopic superpositions, for example, in the context of Refs.~\cite{Schrodinger1935,Leggett1980,Leggett2002}.

\subsection{Distinguishability by classical photon number measurement}
\label{dcn}

Sekatski \textit{et al.}'s measure for optical states is determined by how fuzzy (or how classical) a single-shot photon number measurement can be to distinguish two constituent states of a superposition \cite{Sekatski2014}.
A pointer system is initially assumed to be in a Gaussian position distribution $p_i(x)$ with variance $\sigma^2$.
When a quantum state $\ket{S}$ interacts with the pointer, the final distribution of the pointer becomes \cite{Sekatski2014}
\begin{equation}
p_S(x)=\Tr \left[p_i(x+\hat{a}^\dagger \hat{a})\pjct{S}{S}\right].
\end{equation}
For instance, if $\ket{S}$ is an $n$ photon number state $\ket{n}$, the resultant distribution is exactly shifted by $-n$ from the original one.
When $\sigma$ become larger, the resolution of the detector degrades and the detector is considered to be more ``classical.'' 
The probability of a correct discrimination between two constituent states, $\ket{A}$ and $\ket{B}$, is
\begin{equation}
P^\sigma\left[\ket{A},\ket{B}\right]=\frac{1}{2}\left(1+D\left[p_A^\sigma(x),p_B^\sigma(x)\right]\right),
\end{equation}
where $D\left[p_A^\sigma(x),p_B^\sigma(x)\right]=2^{-1}\int dx \vert p_A^\sigma(x)-p_B^\sigma(x)\vert$ is the trace distance between the outcome distributions $p_A^\sigma(x)$ and $p_B^\sigma(x)$.
The size of a superposition $\ket{A}+\ket{B}$ is then defined by the maximum tolerable $\sigma$ for $P^\sigma\left[\ket{A},\ket{B}\right]$ to reach a certain reference value $P_g$.

A straightforward example discussed in Ref.~\cite{Sekatski2014} is a superposition of the vacuum  and a coherent state $\ket{0}+\ket{\alpha}$  (unnormalized). The measure is obtained as $\abs{\alpha}^2-2(\tn{erf}^{-1}(2P_g-1))^2$ for large $\alpha$, which is proportional to the average photon number $\abs{\alpha}^2$ for a fixed $P_g$.
A more controversial example is an entangled state generated by applying the displacement operation $D(\alpha)=\exp(\alpha \hat{a}^\dagger-\alpha^* \hat{a})$ on single-photon entanglement \cite{Magda2011}
\begin{equation}
\ket{\psi'_D} =  \frac{1}{\sqrt{2}}\left(\ket{1}_A D_B(\alpha) \ket{0}_B + \ket{0}_A D_B(\alpha)\ket{1}_B \right)
\label{eq:GisinState}
\end{equation}   
This state can be rewritten as  $D_B(\alpha)(\ket{+}_A\ket{-}_B-\ket{-}_A\ket{+}_B)$ 
with $\ket{\pm}=(|0\rangle\pm|1\rangle)/\sqrt{2}$ and $D(\alpha)\ket{\pm}$ can be considered as the constituent states of entanglement. 
The measure is then found to be
\begin{equation}
2\alpha {\rm Erf}^{-1}(2P_g-1)\sqrt{\frac{1}{\pi(2P_g-1)^2}-2},
\end{equation}
which is proportional to $\alpha$ indicating that such a state is macroscopically quantum.

There arises a nontrivial question. The coherent states  are considered 
to be the most classical states among all pure states \cite{Sch1926,Mandel1986}.
A coherent state $|\alpha\rangle$ can be rewritten as 
 $|\alpha\rangle=(D(\alpha)|+\rangle+D(\alpha)|-\rangle)/\sqrt{2}$, and 
  $D(\alpha)|+\rangle$ and $D(\alpha)|-\rangle$ are macroscopically distinguishable  according to this approach.
We then have to conclude that a coherent state, as far as its amplitude is large enough, is a macroscopic superposition, which would probably be unacceptable to many researchers in the  community.
It could also be controversial whether merely a local displacement operation can generate genuine macroscopic quantumness.

\section{Implementations of macroscopic quantum states of light}

Various attempts have been made for generating and detecting macroscopic/mesoscopic quantum superpositions using atomic and molecular systems \cite{MonroeCat,C60},  optical setups \cite{Brune96,haroche2008,Ourjoumtsev,Gao,Afek}, superconducting circuits \cite{SQUID1,SQUID2}, and other mechanical systems \cite{mechanical-review}. 
Quantum optics has provided a 
testbed for such experimental implementations in  free-traveling light fields.
For example, there has been remarkable progress in generating superpositions of coherent states (SCSs)
and entangled coherent states (ECSs).
Another example is entanglement between macroscopic and microscopic systems
in several different forms \cite{Martini2008,Jeong2013,Laurat2013} in line with Schr\"odinger's original paradox
\cite{Schrodinger1935}. 
There are yet other examples for generating macroscopic quantum states
such as squeezed macro-macro entanglement 
 \cite{Iskhakov2011pra,Iskhakov2011prl,Iskhakov2012,Kanseri2013}, GHZ-type entanglement \cite{GHZ1,GHZ2,GHZ3,GHZ4} and NOON states \cite{NOON1, NOON2, NOON3, NOON4} that are beyond the scope of this paper.
In this section, we review several of such states and their generation schemes that have been suggested 
and experimentally performed. 
We also discuss their degrees as macroscopic quantum superpositions or entanglement based on some of the measures discussed in the previous section.

\subsection{Superpositions and entanglement of coherent states}

Let us first consider an SCS in the form of 
\begin{equation}
|{\rm SCS}\rangle={\cal N}_\varphi(|\alpha\rangle+e^{i\varphi}|-\alpha\rangle)
\end{equation}
where $|\pm\alpha\rangle$ are coherent states of amplitudes $\pm\alpha$,
$\varphi$ is a real relative phase factor, and ${\cal N}_\varphi$ is the normalization factor.
The SCSs are often referred to as ``Schr\"odinger cat states,''
provided that $\alpha$ is reasonably large, probably due to the following two reasons.
First, the coherent states are known as classical states due to several reasons \cite{Sch1926,Mandel1986,Zurek1993}.
A classic criterion of nonclassicality for a quantum state is whether its $P$-function
\cite{G1963,S1963}
  is well defined \cite{Mandel1986}.
  The coherent states are the only kind of pure states that have well-defined $P$-functions \cite{Mandel1986}.
They are also robust against decoherence as ``pointer states'' \cite{Zurek1993}
and the closest analogy of classical point particles in the quantum phase space \cite{Sch1926}. 
Second, the two coherent states, $|\pm\alpha\rangle$, are macroscopically distinguishable 
when $\alpha$ is sufficiently large;
they can be efficiently discriminated using homodyne detection with a limited efficiency.
Therefore, an SCS generates quantum interferences between these
two ``classical'' but  ``macroscopically distinct''  states as an analogy of Schr\"odinger's paradox.
Schleich {\it et al.}
studied nonclassical properties of SCSs such as sub-Poissonian and oscillatory photon statistics 
\cite{WScat}.
One may also consider entangled coherent states (ECSs) and one of the simplest forms of such states is
\begin{equation}
|{\rm ECS}\rangle= {\cal N}'_\varphi(|\alpha\rangle|\alpha\rangle+e^{i\varphi}|-\alpha\rangle|-\alpha\rangle)
\end{equation}
with normalization factor ${\cal N}'_\varphi$.

Several measures of macroscopic quantum superpositions confirm that SCSs and ECSs are clearly macroscopically quantum.
The values of measure $\cal I$ 
for an SCS and an ECS are found to be
$\mathcal{I}(\ket{\tn{SCS}})=\abs{\alpha}^2(1- e^{-2\abs{\alpha}^2})/(1+ e^{-2\abs{\alpha}^2})$
and $\mathcal{I}(\ket{\tn{ECS}})=2\abs{\alpha}^2(1- e^{-4\abs{\alpha}^2})/(1+ e^{-4\abs{\alpha}^2})$
for $\phi=0$.
In the case of $\alpha\gg1$, they are $\mathcal{I}(\ket{\tn{SCS}})\approx\abs{\alpha}^2$ and 
 $\mathcal{I}(\ket{\tn{ECS}})\approx2\abs{\alpha}^2$.
As another evidence, we numerically obtain the maximum value of $S$  for an SCS to violate Cavalcanti and Reid's inequality~(\ref{eq:Cavalcanti}) \cite{Cavalcanti} using homodyne detection is $S_{\max}\approx2\abs{\alpha}$ which corresponds to the distance between the two peaks of the superposition in the phase space.

\subsubsection{Schemes based on Kerr nonlinear interactions}

Yurke and Stoler found that a coherent state in a Kerr nonlinear medium evolves to
an SCS in the form of $(|\alpha\rangle+i|-\alpha\rangle)/\sqrt{2}$ after a certain interaction time  \cite{YS1986},
based on the calculations for an anharmonic oscillator coupled to a zero-temperature heat bath by Milburn and Holmes \cite{Milburn1986,MH1986}.
They explicitly referred it to as a  ``quantum mechanical superposition of macroscopically distinguishable states''  \cite{YS1986}, quoting Einstein and Sch\"odinger \cite{Schrodinger1935,Einstein1950}.
Mecozzi and Tombesi  showed that  a type of two-mode ECSs can be generated
using a two-mode nonlinear interaction \cite{Tom1987}.
Sanders proposed a scheme to generate ECSs using
a Mach-Zehnder interferometer with a single-mode Kerr nonlinearity
\cite{Sanders1992a}
and later provided a comprehensive review on this kind of states \cite{SandersReview}.
Gerry suggested using  a cross-Kerr nonlinearity in a Mach-Zehnder inteferometer setup with two photodetectors
to generate SCSs and ECSs \cite{Gerry1999}.
Similar schemes using cross-Kerr effects and beam-splitter interferences were independently developed by D'Ariano {\it et al.} \cite{Paris99a} and Howell and Yeazell \cite{Howell2000}. 
Paris devised a two-step scheme that generates a small SCS using a Kerr nonlinear effect and then amplifies it using $\chi^{(2)}$ nonlinearity \cite{Paris99}.
Paternostro {\it et al.} studied a method to generate SCSs and ECSs 
via cross phase modulation in a double electromagnetically induced transparency (EIT) regime \cite{Paternostro02}.
However, all these approaches have  remained as theoretical proposals yet.
The required levels of nonlinearities are extremely demanding even though there are developing
techniques such as EIT to obtain giant Kerr effects \cite{Hau1999}.

There have been trials to reduce the required levels of nonlinearities.
Jeong {\it et al.} found that it is possible to use a relatively weak single-mode Kerr nonlinearity 
together with a conditioning homodyne measurement to generate SCSs \cite{Jeong2004}.
The idea of using a weak cross-Kerr interaction has been discussed in the context of the development of the Bell-state detection \cite{BM2005} and quantum computation \cite{NM2004,MK2005}.
It was explicitly shown \cite{Jeong2005} that a cross-Kerr nonlinearity can be used to generate SCSs and ECSs
and this approach can overcome decoherence due to photon losses.
There exist nontrivial problems to overcome even at a theoretical level 
in order to utilize the approach based on a weak cross-Kerr nonlinearity 
\cite{anti1,anti2,anti3,BingHe2010}. Recently, He {\it et al.} showed that approximately 
ideal cross-Kerr effects can be obtained using relatively weak interactions between photon pulses in atomic ensembles \cite{BingHe2014}.

\subsubsection{Non-deterministic schemes}
\label{subsec:SCSgeneration}
Non-deterministic methods based on linear optics and conditioning measurements have been investigated  \cite{Song1990,Dakna1997,Montina1998,MartiniCat1998,Lund2004,JeongLund2005,
Paris2005,EP2005,ParisLaser2006,
Lance2006,Lance2006b,Marek2008} since initial attempts by several authors \cite{Song1990,Dakna1997,Montina1998,MartiniCat1998}. 
Lund {\it et al.} showed that an SCS with a small amplitude as $\alpha<1.2$ can be well approximated ($F>0.99$) by applying the squeezing operation on a single photon \cite{Lund2004}. It can be achieved by subtracting a single photon from a squeezed state and a number of experiments have been performed along this line \cite{kt1,kt2,kt3,kt4}.
However, this type of experiment cannot generate two separate peaks in the phase space with interferences between them because the resulting state is simply a squeezed form of the single photon. 
Ourjoumtsev {\it et al.} attempted a different method: a number state is divided by a 50:50 beam splitter and conditioned by a homodyne detection to generate an SCS \cite{Ourjoumtsev}. The experiment with a number state of $n=2$ has resulted in an approximate SCS with $\alpha\approx1.6$ with interferences between two clearly separate peaks in the phase space \cite{Ourjoumtsev}.
It can be used to generate arbitrarily large SCSs if large number states are available \cite{Ourjoumtsev}.
Glancy and  de Vasconcelos provided a comprehensive review on various methods for producing SCSs in optical systems in 2008 \cite{Glancy2008}. 
Marek {\it et al.} showed that multiple photon subtractions result in large squeezed SCSs
with extremely high fidelities \cite{Marek2008}.
In this way, demonstrations of subtracting two or more photons on a squeezed vacuum state have been performed to generate SCSs \cite{Sasaki2008a,Sasaki2008b,Gerrits2010}.

\begin{figure}[t]
\centering
\includegraphics[width=0.8\linewidth]{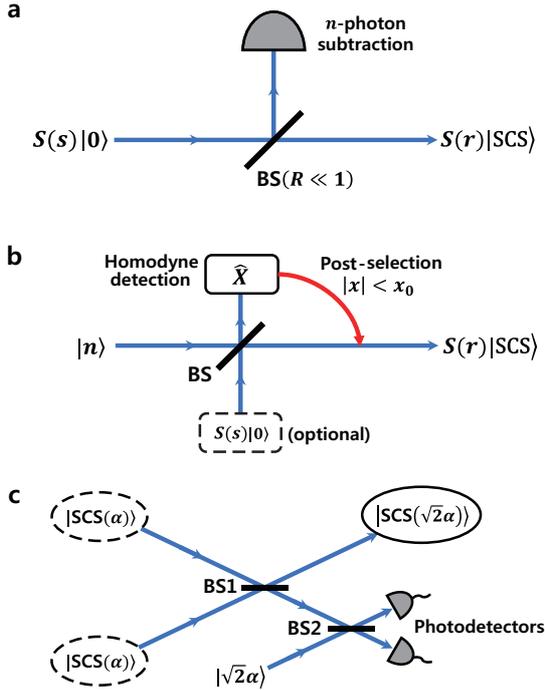}
\caption{
Non-deterministic generation methods for SCSs and their amplification scheme \cite{kt2,Sasaki2008a,Gerrits2010}.
(a) $N$-photon subtraction on a squeezed vacuum $\hat{S}(r)|0\rangle$ using a beam splitter (BS) with low reflectivity ($R\ll1$) and a photodetector. 
(b) SCS generation scheme using homodyne detection \cite{Ourjoumtsev,Lance2006}.
  Homodyne conditioning for measurements outcomes $|x|<x_0$  using a 50:50 beam splitter (BS) and homodyne detection $\hat X$ on a photon number state $|n\rangle$.
The additional (optional) squeezed vacuum $S(s)|0\rangle$ may be used to control the degree of squeezing $r$ of the the output SCS.
(c) The SCS amplification scheme \cite{Lund2004}. Two small SCSs with amplitudes $\alpha$ may be used to conditionally generate an SCS of a larger amplitude $\sqrt{2}\alpha$ using two 50:50 beam splitters (BS1 and BS2), an auxiliary coherent state $|\sqrt{2}\alpha\rangle$ and two {\it imperfect} on/off photodetectors.
}\label{fig-scs}
\end{figure}

Many of the non-deterministic yet feasible schemes are categorized into two major approaches. One is based on the squeezing operation and photon subtractions \cite{Dakna1997,Lund2004,JeongLund2005,Marek2008,Sasaki2008a,Sasaki2008b,Gerrits2010} and the other is based on the number state generation and homodyne detection \cite{Ourjoumtsev,Lance2006,Lance2006b}. Figures~\ref{fig-scs}(a) and 1(b) present schematics of the two approaches. In fact, either the multiple photon subtraction from a squeezed vacuum \cite{Marek2008} or homodyning on the one part of the number state divided by a beam splitter  \cite{Ourjoumtsev} yields a {\it squeezed} SCS, i.e. $S(r)|{\rm SCS}\rangle$ where $S(r)$ is the squeezing operator with the squeezing parameter $r$ as depicted in Fig.~\ref{fig-scs}.
In principle, as far as one could perform the photon number subtraction of many photons \cite{Marek2008} or generate a large number state \cite{Ourjoumtsev}, a squeezed SCS of a large amplitude and high fidelity can be obtained.
In order to generate a normal SCS (without squeezing) with high fidelity $F>0.9999$, in addition to the scheme in Ref.~\cite{Ourjoumtsev}, one may use an additional squeezed vacuum as shown in Fig.~\ref{fig-scs}(b) \cite{Lance2006,Lance2006b}.
A generation scheme for arbitrary SCSs with unbalanced ratios was suggested \cite{cat-arb-1} and experimentally demonstrated \cite{cat-arb-2}.
It is worth noting that large SCSs may be obtained in a non-deterministic way out of small SCSs using realistic on/off detectors and an auxiliary coherent state as shown in Fig.~\ref{fig-scs}(c) \cite{Lund2004,JeongLund2005}, and there exists an alternative amplification scheme using homodyne detection \cite{Ami}.

In general, the SCSs and ECSs are sensitive to a lossy environment and detection inefficiency as their amplitudes become large \cite{KimBuzek1992}. 
Theoretical attempts were made using the squeezing operation to make the SCSs more robust against losses \cite{KLB1993,ParisSqueezedCat} or to amplify them
\cite{RadimAmp1,RadimAmp2}.
Schemes for entanglement purification for mixed ECSs \cite{puri,puri0} and concentration for pure ECSs \cite{puri} were investigated using Bell-state measurements \cite{puri}.
Ourjoumtsev {\it et al}. devised and experimentally demonstrated a method to generate an ECS in a remote way using two SCSs and two photodetectors \cite{Ourj-ECS}.
Lund {\it et al}. suggested another scheme for the same purpose that is made to be more robust against detection inefficiency using an auxiliary coherent state \cite{Lund2013}.
Proposals to distribute ECSs using the quantum repeater protocol were also suggested \cite{Brask, Repeater}.
Superpositions and entanglement of multiple numbers of coherent states have been theoretically studied \cite{KL1993,Wang01,Enk03,Enk05}. Lee {\it et al.} experimentally demonstrated quantum teleportation of a single-photon-subtracted squeezed vacuum (i.e., an approximate SCS) using the continuous variable teleportation protocol \cite{Furusawa2011}.
Generalization of SCSs and ECSs to highly mixed forms has been studied for testing quantum theory in an even more ``classical'' limit \cite{JeongRalph2006,JeongRalph2007}.

\subsubsection{Applications}

The wide scope of applications using SCSs and ECSs includes quantum teleportation \cite{Wang01,Enk01,JKL01,puri,BaAn2003}, quantum computation \cite{Cochrane1999,Jeong02,Ralph03,LundRalphPRL,Marek2010,MR_NJP,KIAScom10}, precision measurements \cite{Gerry01,Gerry02,Gerry03,WeakForce,Jaewoo1,Hirota-arx11,Jaewoo2}, Bell-type inequality tests \cite{Mann1995,Filip2000,Derek,jeongsonkim,WegBell2003,Magda,sole2008,LJsq2009,
Gerry2009pra,Jeong2009,Lim2012,Kirby2013,Kirby2014},
 Leggett-type inequality tests \cite{LG_J2010,LG_J2011},  quantum contextuality tests \cite{contex} and quantum steering \cite{steering-ecs}. In particular, quantum teleportation and computation schemes using SCSs and ECSs are considered as a strong candidate for the optical implementation of quantum computation in terms of the resource requirement and loss tolerance \cite{RP2009}. A proof-of-principle experiment of this type of teleportation scheme was performed by Neergaard-Nielsen {\it et al}. \cite{tele-amp}.

\subsection{Towards microscopic-macroscopic entanglement }

There have been attempts to generate entanglement between microscopic and macroscopic (or quantum and classical) states of light by using several different methods \cite{Magda2011,Martini2008,DeMartiniQIOPA,DeMartiniPLA, Sekatsky2010,Ghobadi2013,Gisin2013,Lvovsky2013,Jeong2013,Laurat2013}. 
Implementations of such micro-macro entanglement are of special interest in relation to Schr\"odinger's cat paradox where a cat as a macroscopic classical object and an atom as a microscopic quantum system are entangled \cite{Schrodinger1935}.
We limit our discussions to several examples using free-traveling optical systems mainly from a theoretical point of view, while there are other examples
such as atom-field entanglement \cite{Brune96,haroche2008}. 

\begin{figure}[t]
\centering
\includegraphics[width=0.8\linewidth]{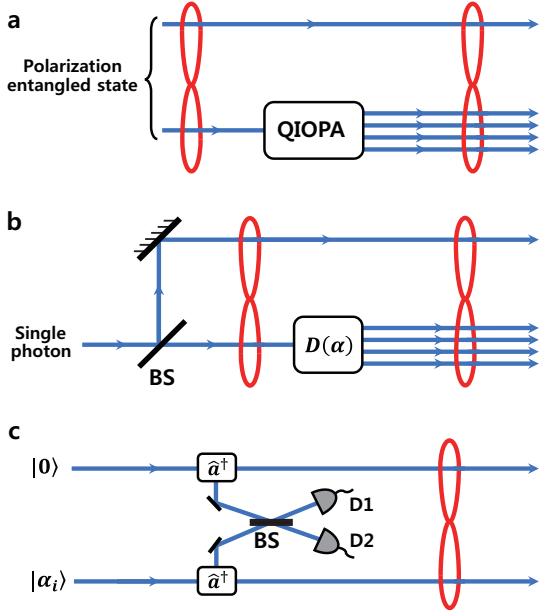}
\caption{Schematics for generating micro-macro entanglement.
(a) A part of two-photon polarization entanglement is amplified using quantum injected optical parametric amplification (QIOPA) \cite{Martini2008}. 
(b) A part of single-photon entanglement is displaced to have a large photon number
\cite{Gisin2013,Lvovsky2013}.
(c) Entanglement between a single photon and a coherent state is generated using a coherent superposition of two distinct operations \cite{Jeong2013}.
}
\label{Fig-micro-macro}
\end{figure}

\subsubsection{Amplifying microscopic entanglement by nonlinear amplifier}
\label{sec:amp-mm}

De Martini {\it et al}. suggested  \cite{Martini2008,DeMartiniQIOPA,DeMartiniPLA} and experimentally demonstrated   \cite{Martini2008} a scheme depicted in Fig.~\ref{Fig-micro-macro}a to amplify a local part of microscopic entanglement using quantum injected optical parametric amplification (QIOPA)  \cite{Martini2008} \cite{DeMartiniQIOPA, DeMartiniPLA}.
The microscopic entanglement is supposed to be a polarization entangled state,
$\left(\ket{R}_{A} \ket{L}_B - \ket{L}_A \ket{R}_B \right)/\sqrt{2}$,
where 
$\ket{R}_A$ ($\ket{L}_B$) is the left (right) circularly polarization state 
for mode $A$ ($B$).
One of the modes, say $B$, is amplified to be ``macroscopic'' using QIOPA
\cite{DeMartiniQIOPA, DeMartiniPLA}.
The states $\ket{R}$ and $\ket{L}$ then evolve to two orthogonal states with different photon number distributions as
\begin{align}
\ket{R}\rightarrow \ket{\Phi^R}=\sum_{i,j=0}^\infty \Delta_{ij} \ket{2i+1;R}\ket{2j;L} \nonumber \\
\ket{L}\rightarrow \ket{\Phi^L}=\sum_{i,j=0}^\infty \Delta_{ij} \ket{2i+1;L}\ket{2j;R}
\label{eq:process}
\end{align}
where $\ket{k;R}$ ($\ket{k;L}$) is the photon number state with $k$ photons and 
the right (left) circularly polarization, and  $\Delta_{ij}$ are real coefficients\footnote{ $\Delta_{ij}= \left(\cosh{g}\right)^{-2} \left(-\frac{\tanh{g}}{2} \right)^i \left( \frac{\tanh{g}}{2} \right)^j \frac{\sqrt{(2i+1)!(2j)!}}{i!j!}$} with the amplification constant $g$.
The amplification results in entanglement between microscopic and macroscopic states:
\begin{equation}
|\Phi_g\rangle= \frac{1}{\sqrt{2}} \left( \ket{R}_A \ket{\Phi^L}_B  - \ket{L}_A \ket{\Phi^R}_B \right).
\label{MartiniState}
\end{equation}

It is straightforward to verify that state $|\Phi_g\rangle$ has the maximum value of measure $\cal I$ \cite{LeeJeong2011} as
${\cal I}(|\Psi_g\rangle)=1+\sum_{i,j=0}^\infty \Delta_{ij}^2 (2i+2j+1)$ that corresponds to its 
average photon number.
This result is implied in the fact that the QIOPA causes $|R\rangle$ and $|L\rangle$ to be macroscopically quantum
as seen in Eq.~(\ref{eq:process}).
 In fact,  $\ket{\Phi^R}$ and $\ket{\Phi^L}$ themselves have the maximum values of $\cal I$.
However, it is a separate question whether 
$\ket{\Phi^R}$ and $\ket{\Phi^L}$ are truly macroscopically distinct in this type of entanglement.
In Ref.~\cite{Martini2008}, the two states $\ket{\Phi^R}$ and $\ket{\Phi^L}$ are considered to be macroscopically distinct in the sense that their average photon numbers for the $R$-polarization are $\sinh^2{g}$ and $3\sinh^2{g}+1$, respectively.
However, 
as we discussed in Sec.~\ref{dcn}, this type of approach leads to a problem
that becomes more noticeable when it is applied to a single-mode superposition.
 Once again, the example of a strong coherent state, $|\alpha\rangle=
D(\alpha)|+\rangle+D(\alpha)|-\rangle$ with $\alpha\gg1$, may be considered for a comparison.
It can be simply shown that 
the difference between the average photon numbers of constituent states $D(\alpha)|+\rangle$ and $D(\alpha)|-\rangle$ is $2|\alpha|$, i.e., it monotonically increases with $\alpha$ although a coherent state would not be regarded as a macroscopic superposition. In spite of this ambiguity, a strong nonclassical feature of
state $|\Phi_g\rangle$ seems evident based on the high value of $\cal I$.

A perfect discrimination of  $\ket{\Phi^R}$ and $\ket{\Phi^L}$  by a single-shot measurement requires a photon-number resolving detector for a parity measurement.
In the real experiment \cite{Martini2008}, an alternative  procedure called the ``orthogonality filter'' was used where $\ket{\Phi^R}$ and $\ket{\Phi^L}$ are discriminated by field intensity measures for each polarization in a non-deterministic and approximate way. 
There have been considerable debates and discussions \cite{Sekatski2009,Sekatsky2010, Raeisi2011,Raeisi2012,MartiniDiscussion}  over whether the state generated in  Ref.~\cite{Martini2008} possesses genuine entanglement and how to verify this type of entanglement more clearly.

\subsubsection{Amplifying microscopic entanglement by local displacement operations}

Sekatski {\it et al.} proposed a scheme that uses the displacement operation for an amplification process to generate macroscopic entanglement \cite{Magda2011}.
The state is generated by locally displacing each mode of single-photon entanglement as
$|\psi''_D\rangle=
D_A(\alpha)D_B(\alpha)(\ket{1}_A\ket{0}_B-\ket{0}_A\ket{1}_B)/\sqrt{2}$.
They claim $D(\alpha)\ket{0}$ and $D(\alpha)\ket{1}$ are macroscopically distinguishable since they exhibit 3 times different photon number variances \cite{Magda2011}.
Bruno {\it et~al.} \cite{Gisin2013} and Lvovsky {\it et~al.} \cite{Lvovsky2013} experimentally realized a simpler variant $\ket{\psi'_D}$ in Eq.~(\ref{eq:GisinState}), where only one of the modes is displaced
for generating micro-macro entanglement (see Fig.~\ref{Fig-micro-macro}b).

In Refs.~\cite{Gisin2013,Lvovsky2013}, it was shown that a single shot photon number measurement can distinguish between $D(\alpha)\ket{0}$ and $D(\alpha)\ket{1}$ in the macroscopic part ($B$) of state (\ref{eq:GisinState}) with 74\% probability for $\alpha\gg1$.
It can be improved up to 90\% by changing the basis of the initial microscopic state into $\ket{\pm}=(\ket{0}\pm\ket{1})/\sqrt{2}$.
However, we already discussed in Secs.~\ref{dcn} and \ref{sec:amp-mm} that this type of approach does not seem very convincing.
The value of measure $\cal I$ for $\ket{\psi_D'}$, as well as $\ket{\psi_D''}$,  is exactly the same to that of microscopic single-photon entanglement $(|0\rangle|1\rangle-|1\rangle|0\rangle)/\sqrt{2}$.
In fact, measure $\cal I$ is invariant under the displacement operation that does not change the structure of phase space distributions of any states \cite{LeeJeong2011}. The same principle is applied to Cavalcanti and Reid's inequality \cite{Cavalcanti}
so that neither $\ket{\psi_D'}$ nor $\ket{\psi_D''}$ is found macroscopically quantum.

In the experiments \cite{Gisin2013,Lvovsky2013}, entanglement was detected after applying the ``reverse'' displacement operation, $D_B(-\alpha)$, to state (\ref{eq:GisinState}). Bruno {\it et~al.} observed  entanglement with more than 500 photons using photodetectors \cite{Gisin2013}.
Lvovsky {\it et~al.} achieved the displacement operations with $|\alpha|^2 \approx 1.6 \times 10^8$ and entanglement was observed using homodyne tomography \cite{Lvovsky2013}.

Ghobadi {\it et al.}~\cite{Ghobadi2013} investigated an alternative proposal to create micro-macro entanglement by applying the squeezing operation $S(r)$ to one mode of single-photon entanglement as
$\ket{\psi_S} =  S_B(r) \left(\ket{1}_A  \ket{0}_B + \ket{0}_A \ket{1}_B \right)/\sqrt{2}$.
It was shown that two states $S(r) \ket{0}$ and $S(r) \ket{1}$ can be discriminated by the mean photon number contained in each state which are 3 times different for large enough squeezing.
The squeezing operation changes the structure of the phase space distribution and  the degree of macroscopic quantumness  
$\mathcal{I}(\ket{\psi_S})= 2\sinh^2 r +1$  increases as $r$ becomes larger while the tricky issue explained above on macroscopic distinctness between  $S(r) \ket{0}$ and $S(r) \ket{1}$ remains. This state can be categorized into the same type of state with the one in Eq.~(\ref{MartiniState}).

\subsubsection{Generating hybrid entanglement by photon addition or subtraction}
Recently, hybrid entanglement between particle-like (or quantum) and wave-like (or classical) states was experimentally demonstrated by Jeong {\it et al.} \cite{Jeong2013} and Morin {\it et al.} \cite{Laurat2013}.  A form of such state is
\begin{equation}
\ket{\Psi_\alpha} = \frac{1}{\sqrt{2}} \left( \ket{0} \ket{\alpha} + \ket{1} \ket{-\alpha} \right).
\label{Hybrid}
\end{equation}
It takes the maximum value of $\cal I$ for $\alpha\gg1$ as
$\mathcal{I}(\ket{\Psi_\alpha})\approx|\alpha|^2+1/2$.
Jeong {\it et al.}'s scheme uses an idea of superposing two distinct quantum operations.
Initially, a coherent state $|\alpha_i\rangle$ and the vacuum are prepared with two photon addition devices
\cite{ZavattaAddition}
and a beam splitter with appropriate ratio as shown in Fig.~\ref{Fig-micro-macro}(c).
When only of the two detectors D1 and D2 clicks, one does not know which mode the single photon was added to.
The resulting state is then superposition as $\hat{a}_A^\dagger \ket{0}_A \ket{\alpha_i}_B +\ket{0}_A\hat{a}_B^\dagger  \ket{\alpha_i}_B $ (unnormalized) that can be made to be an approximate state of $\ket{\Psi_\alpha}$ by applying an appropriate displacement operation.
However, hybrid entanglement with only small values of $\alpha$ can be obtained in this way.
Experimentally, a fidelity of $F\approx 0.76$  and entanglement of negativity ${\cal N} \approx 0.45$ were shown 
with $\alpha\approx0.31$ \cite{Jeong2013}.
More sophisticated methods such as the tele-amplification method \cite{tele-amp} need be used to obtain large values of $\alpha$ for  such hybrid states \cite{Jeong2013}.

Morin {\it et~al.}'s experiment \cite{Laurat2013} is based on a similar type of idea but using the photon subtraction with a beam splitter and a photodetector, a two-mode squeezed state with a very low gain, and a single-mode squeezed state in order to generate a slightly different type of target state such as
$( \ket{+} \ket{\alpha} + e^{i \varphi} \ket{-}  \ket{-\alpha})/\sqrt{2}$.
Entanglement of negativity ${\cal N} \approx 0.7$ and fidelity  $F\approx 0.77$ were observed with
amplitude $\alpha \approx 0.9$ \cite{Laurat2013}.
This type of idea is also found in Andersen and Neergaard-Nielsen's previous proposal where
a single photon, a single-mode squeezed state and a detector with a beam splitter for the photon subtraction are required \cite{Andersen2013}. The target states of all these schemes \cite{Jeong2013,Laurat2013,Andersen2013} show strong properties as macroscopic quantum states, i.e., the maximum values of $\cal I$ for $\alpha\gg1$.
Kreis and van Loock investigated how to classify and quantify various types of hybrid entanglement
between discrete and continuous variable states \cite{KreisLoock2012}.

\subsubsection{Applications}

Entanglement between microscopic and macroscopic states is closely related to Schr{\"o}dinger's  Gedankenexperiment \cite{Schrodinger1935}.
Micro-macro (or hybrid-type) entangled states are 
useful for loophole-free Bell-type inequality tests \cite{Spagnolo2011, Stobinska2011,Kwon2013,Stobinska2014,MartiniarXiv2008},
quantum information processing \cite{Furusawa2011b,Loock2011,Park2012,Lee2013,Rigas2006, Wittmann2010},
and exploring quantum gravity \cite{Simon2004}.
In particular, optical hybrid states benefit from both their discrete- and continuous-variable features in such a way that they are useful resources for quantum teleportation \cite{Park2012,Lee2013}, quantum computation \cite{Lee2013}, and quantum key distribution \cite{Rigas2006, Wittmann2010}.
Sheng {\it et al.} proposed an entanglement purification scheme for hybrid entanglement using
linear optics elements and photon number measurements \cite{Sheng2013}.

\section{Remarks}
We have reviewed and discussed proposed measures of macroscopic quantumness together with several attempts for implementing macroscopic quantum states using optical fields.
In line with our discussions, three \cite{LeeJeong2011,Frowis2012NJP,Nim2013} among the discussed measures seem to be a little more general than the others in the sense that they are applicable to various kinds of quantum states and serve as unambiguous quantifiers.
Those measures are particularly suitable for light fields \cite{LeeJeong2011}, spin systems \cite{Frowis2012NJP}, and mechanical systems \cite{Nim2013}, respectively. It  seems that the former two \cite{LeeJeong2011,Frowis2012NJP} stick to the idea of genuine macroscopic quantum superpositions \cite{Schrodinger1935,Leggett1980,Leggett2002}, i.e., superpositions of macroscopically distinct states or genuine macroscopic quantum effects over mere accumulation of microscopic quantum effects. On the other hand, the latter \cite{Nim2013} seems to approach macroscopic quantumness in a broader way.

It would be interesting to find out relations between measures for different physical systems from an inclusive point of view. 
Very recently, Fr\"owis \textit{et al.} investigated macroscopic quantumness of several optical states using measures for spin systems \cite{FrowisLinking}.
They 
mapped a photonic state into a spin state using an ideal interaction model
and found that some different measures show strong mathematical connections and give a similar classification of macroscopic quantum states.

There have been various attempts to generate macroscopic quantum states using light fields.
The arbitrariness of the decomposition for a quantum state leads to a question of whether certain quantum states \cite{Gisin2013,Lvovsky2013} are truly macroscopically quantum, at a theoretical level, even though they seem so with certain criteria
\cite{Sekatski2014}.
We exemplified the case where even a coherent state, $|\alpha\rangle$, is categorized into a macroscopic superposition under a certain decomposition using this type of approach \cite{Sekatski2014}. Perhaps, this suggests that more stringent considerations are necessary 
for classifications of macroscopic quantum states.

\section*{Acknowledgements}

This work was supported by the National Research
Foundation of Korea (NRF) grant funded by the Korea
government (MSIP) (No. 2010-0018295) and the Center for Theoretical
Physics at Seoul National University.

~

\end{document}